\documentclass{article}

\usepackage[alldates=year, block=space, style=apa]{biblatex}
\usepackage[margin=1in]{geometry}
\usepackage[colorlinks, allcolors=blue]{hyperref}

\usepackage{amsfonts, amsmath, authblk, float, graphicx, tikz}
\usetikzlibrary{decorations.pathmorphing}

\title{Swinging, Fast and Slow:\\Interpreting variation in baseball swing tracking metrics}
\author[1*]{Scott Powers}
\author[2*]{Ronald Yurko}
\affil[1]{Department of Sport Management, Rice University}
\affil[2]{Department of Statistics \& Data Science, Carnegie Mellon University}
\date{July 1, 2025}

\begin{document}

  \maketitle

  \def\thefootnote{*}\footnotetext{Both authors contributed equally to this work.}
	
  \begin{abstract}
    In 2024, Major League Baseball released new bat tracking data, reporting swing-by-swing bat speed and swing length measured at the point of contact. While exciting, the data present challenges for their interpretation. The timing of the batter's swing relative to the pitch determines the point of measurement relative to the full swing path. The relationship between swing metrics and swing outcomes is confounded by the batter's pitch recognition. We introduce a framework for interpreting bat tracking data in which we first estimate the batter's intention conditional on ball-strike count and pitch location using a Bayesian hierarchical skew-normal model with random intercept and random slopes for batter. This yields batter-specific effects of count on swing metrics, which we leverage via instrumental variables regression to estimate causal effects of bat speed and swing length on contact and power outcomes. Finally, we valuate the tradeoff between contact and power due to bat speed by modeling a plate appearance as a Markov chain. We conclude that batters can reduce their strikeout rate by reducing bat speed as strikes increase, but the tradeoff in reduced power approximately counteracts the benefit to the average batter.
  \end{abstract}

  \section{Introduction}
  \label{sec:introduction}

    Baseball has a long history as fertile ground for statistical analysis of rich datasets. For Major League Baseball (MLB), detailed records are publicly available at the play-by-play level going back to 1912 (\cite{retrosheet_retrosheet_2025}). Beginning in 2007, MLB made optical pitch tracking data freely available online
    (\cite{fast_what_2010}), enabling unprecedented insights into player performance (\cite{swartz_quality_2017}). In 2016, MLB made waves again when they updated their ball tracking technology to incorporate Doppler radar and released batted ball tracking data to the public (\cite{arthur_new_2016}). In May 2024, MLB released their most exciting new dataset since batted ball tracking in 2016 by making bat tracking data available at the pitch-by-pitch level (\cite{petriello_everything_2024}).

    The new bat tracking metrics include {\it bat speed}, {\it fast-swing rate}, {\it squared-up rate}, {\it blasts} and {\it swing length}. Of these, only bat speed and swing length represent new information: The other metrics (fast-swing rate, squared-up rate and blasts) are functions of bat speed and other ball tracking data that were already publicly available. Both bat speed and swing length are measured at the moment of bat-ball contact (or at the point nearest to contact, in the case of a swing-and-miss). Bat speed is the linear speed of the ``sweet spot'' of the bat, approximately six inches from the end; and swing length is the distance traveled by the end of the bat from the start of the swing until contact. Critically, bat speed and swing length are available for each swing, not just aggregated at the player level. This opens the door to pitch-level analysis within player, not just analysis of differences between player averages.

\begin{figure}[H]
      \centering
      \includegraphics[width = 0.8\textwidth]{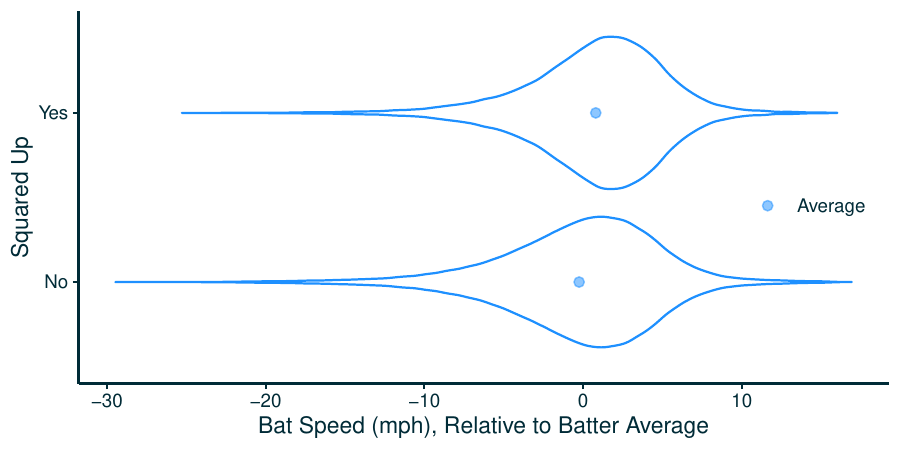}
      \caption{\it Distribution of bat speed relative to batter average by contact quality (squared up or not), across all swings in the dataset. The x-axis represents the difference between the bat speed and the batter's average bat speed. Following MLB's definition, a swing is considered ``squared up'' if the batted ball's exit velocity is at least 80\% of the theoretical maximum (given pitch speed and bat speed), a proxy for solid contact.}
      \label{fig:counterintuitive}
    \end{figure}

    An open research question in baseball is the effect of batter aggressiveness. Conventional coaching wisdom would suggest that the harder a player swings, the less likely they are to make solid contact, but the harder they are likely to hit the ball if they do make solid contact. What is the effect of bat speed on contact? Figure \ref{fig:counterintuitive} illustrates a simple analysis, and the result is counterintuitive. We observe that swings which result in solid contact actually have slightly faster-than-average bat speeds (relative to the batter's own average). This shows that bat speed is positively correlated with making solid contact. Taken at face value, it would seem to suggest that batters should always swing as hard as possible because it does not reduce contact quality and may even increase it.

\begin{figure}
      \centering
      \includegraphics[width = 0.4\textwidth]{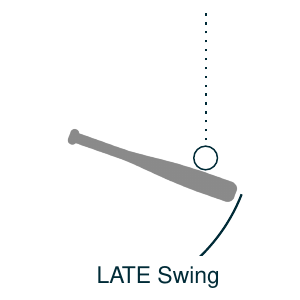}
      \includegraphics[width = 0.4\textwidth]{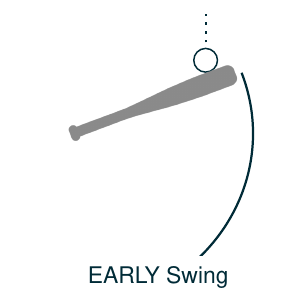}
      \caption{\it A diagram illustrating the effect of swing timing on swing length measurement due to the fact that swing length is measured at the point of contact. Imagining two swings with the exact same mechanics, the left image shows the swing length measurement if the swing is late, and the right image shows the swing length measurement if the swing is early.}
      \label{fig:swing-diagram}
    \end{figure}

    Like the famous problem of the bat and ball whose total cost is \$1.10 (\cite{kahneman_thinking_2011}), this is a problem for which slower, more deliberate thinking is helpful. The dilemma is due to how the data are measured. The measurement occurs at the point of contact, the timing of which is closely related to the outcome of the swing. Figure \ref{fig:swing-diagram} illustrates two hypothetical swings with identical swing mechanics, but because one swing is early and the other is late, the measurements are taken at different points in the swing. Thus, the timing of the swing affects the measured bat speed and swing length. From this perspective, bat speed and swing length are themselves outcomes of the swing, not entirely under the batter's control.

    We hypothesize that the counterintuitive results shown in Figure \ref{fig:counterintuitive} are explained by an unobserved confounder: the extent to which the batter correctly identifies the pitch type early in its trajectory (by lucky guess or by other means---see, e.g., \textcite{elmore_bang_2022}). For example, when the batter prepares for a high fastball and this guess is correct, they are likely to complete their swing with full intent. When the batter preparing for a high fastball is met with a curveball, they need to make a mid-swing adjustment, reducing their bat speed in the process. The latter scenario is less likely to result in contact than the former.

    Our approach to this problem is to model the {\it intent} of the batter rather than taking observed bat speed and swing length at face value. We control for contact quality, and we model the distribution of swing metrics for each batter {\it when they are successful} to better identify what they are trying to do. We estimate separate Bayesian hierarchical skew-normal models for intended bat speed and intended swing length, and we demonstrate heterogeneity between batters in how they modulate their intended swing metrics as the number of strikes increases. We perform instrumental variables regression to estimate the causal effect of these intended swing metrics on swing outcomes. Finally, we model a plate appearance as a Markov chain to evaluate the tradeoff between contact and power when modulating bat speed and swing length.

    \subsection{Related Work}
    \label{sec:related-work}

      Shortly after MLB released the new bat tracking data, there were several analyses published in online outlets. \textcite{clemens_what_2024} explored correlations on the player level between the new bat tracking metrics and existing performance metrics, finding that the new metrics did not explain much variation in overall performance. This was followed by more careful analyses at the swing level. \textcite{woodward_radar_2024} estimated a generalized additive model (GAM) to predict bat speed (as a percentage of the batter's peak bat speed) based on pitch location and other variables; and then manually created rules based on the residuals of this model to label certain individual swings as ``guess'' swings or ``catch-and-redirect'' swings in an effort to identify batter intent. \textcite{orr_defense_2024} pointed out that swings resulting in pulled fly balls tend to have greater swing lengths because the contact point tends to be further in front of home plate; and then fit a GAM to predict swing length from pitch location and bat speed in an effort to isolate contact depth.

      In the academic literature, we have not found any previous studies analyzing newly available bat tracking data or examining the relationship between swing aggressiveness and batting outcome in baseball. \textcite{nevins_sensitivity_2019} modeled the relationship between bat speed and batted ball speed for fastpitch softball batters. \textcite{orishimo_lower_2024} investigated correlations between kinetic/kinematic features and bat speed, finding that ground force reactions (vertical, anterior/posterior, and resultant) under the lead foot were most correlated with bat speed. In perhaps the most intriguing and closely related study to the present work, \textcite{nakashima_acceptable_2025} collected full-path bat tracking data from 145 swings made by 29 collegiate batters in a laboratory setting and calculated for each swing the acceptable range of timing error for contacting a pitch with the ``sweet spot'' of the bat. In a different sport, \textcite{burton_linear_2015} developed a model for the relationship between volleyball serving agressiveness and service outcomes and calculated optimal aggressiveness based on a server's rates of aces and errors.

      Recent research on statistics in sports has used some of the methodological pieces in the present work. \textcite{judge_exit_2024} presented a skew-normal likelihood for the distribution of a batter's exit velocity in baseball, which motivated our approach in Section \ref{sec:methods-intention}. \textcite{putman_tackling_2025} used instrumental variables regression (as we do in Section \ref{sec:methods-causal}) to model the causal effect of pass/run play calling on offensive play outcomes in American football.

  \section{Data}
  \label{sec:data}

    Our dataset comprises 685,143 pitches thrown during the 2024 MLB regular season. This publicly available dataset was downloaded from Baseball Savant (\cite{mlb_advanced_media_statcast_2024}) using the \texttt{R} package \texttt{sabRmetrics} (\cite{powers_sabrmetrics_2025}). For each pitch, we observe the identities and handedness (left/right) of the batter and pitcher involved and the ball-strike count prior to the pitch. We also observe pitch tracking information, most notably the pitch type (e.g. fastball, curveball, changeup) and the $(x, z)$ location where the pitch crosses the front of home plate. The pitch tracking information is sufficient to recreate an estimate of the pitch's full trajectory as a three-dimensional quadratic function of time $(x(t), y(t), z(t))$ and to recreate the corresponding pitch metrics such as release point, release speed, and vertical/horizontal break. Each pitch has a categorical outcome, which can be one of: Hit by Pitch (HBP), Called Ball, Called Strike, Swinging Strike, Foul Ball, or Fair Ball.

    For pitches on which the batter swings, we observe bat tracking information describing the bat speed and the swing length (detailed in Section \ref{sec:measurement}). For pitches that are hit into play (i.e. the Fair Ball outcome), we observe the initial exit speed and vertical exit angle (relative to the ground) of the ball as it leaves the bat. We also observe each fair batted ball's manually charted hit coordinates (on the field), which serves as a proxy for the horizontal exit angle of the ball off the bat.

    \subsection{Bat Speed and Swing Length Measurement}
    \label{sec:measurement}

      Bat tracking metrics were made possible by improvements to MLB's Hawk-Eye Statcast system in 2023 which featured five high frame rate (300 frames per second) cameras dedicated to tracking movements of the pitcher and the batter (\cite{goldbeck_introducing_2023}). Through computer vision, this system tracks not only player center of mass but full limb orientation (including the bat) throughout the pitching and batting movements, generating a rich kinematic time series of data on each pitch.

      The raw kinematic time series data are not publicly available, but MLB has begun to release some metrics derived from them, starting with bat speed and swing length on every pitch. Bat speed is defined as the linear speed of the ``sweet spot'' of the bat (roughly six inches from the end) at the point of contact with the ball, and swing length is defined as the distance traveled by the end of the bat from the start of the swing until the point of contact (\cite{petriello_everything_2024}). In the event of a swing and miss, both metrics are measured at the point nearest to contact. These metrics represent a tiny fraction of the kinematic time series data, and the process by which they are measured complicates their interpretation on a swing-by-swing basis.

    \subsection{Data Cleaning}

    The initial dataset includes attempted bunts (which are labeled) and attempted {\it check-swings} (i.e., partial swings, not labeled). We aim to limit our analysis to full swings. When reporting aggregate metrics like average bat speed, MLB filters out each batter's bottom 10\% of swings by bat speed, labeling the remaining swings as {\it competitive}. We take a more inclusive approach by filtering out only swings with bat speed below 50 mph (which filters out all bunt attempts). Based on video review, this cutoff does a good job of distinguishing between full and partial swings while only removing $\approx$ 2.5\% of swings.

    We reproduce MLB's derived metric for {\it squared-up} contact as:
    \begin{equation}
        \label{eqn:squared-up}
    \mbox{batted ball speed} > 80\% \times \left\{1.23 \cdot (\mbox{bat speed}) + 0.23 \cdot (\mbox{pitch speed})\right\}.
    \end{equation}
    The right-hand side of Equation (\ref{eqn:squared-up}) is a rough approximation for 80\% of the theoretical maximum batted ball speed given bat speed and pitch speed at contact, with additional assumptions about the mass of bat and ball (\cite{nathan_dynamics_2000}). We observe batted ball speeds as high as 110\% of this rough theoretical maximum.

    \subsection{Building Blocks}

      The present work relies on some models that are not novel, but nevertheless warrant explanation for an audience not steeped in baseball analytics literature.
    
      \subsubsection{Linear Weights}
      \label{sec:linear-weights}

        Linear weights (\cite{thorn_hidden_1984}) are the standard metric for the run value of a plate appearance outcome (strikeout, walk, single, etc.). They are based on the concept of base-out run expectancy, which models the progression of an inning as a Markov chain between base-out states, meaning that the state of the Markov chain is defined by which bases are occupied by runners and by the number of outs. For each state, the run expectancy is defined as the expected number of runs scored from that state until the terminal state (end-of-inning, three outs). The transition probability function is generally calculated using the league-wide empirical frequency of transitions between base-out states across a large, multi-year sample of data.

        The linear weight of a plate appearance outcome is defined as the average change in base-out run expectancy that occurs on that outcome. For example, the linear weight of a walk is the average change in base-out run expectancy across all walks in the data. Some walks may result in greater changes in run expectancy (e.g. with bases loaded and zero outs) or lesser changes in run expectancy (e.g. with bases empty and two outs), but the linear weight does not depend on the context in which the outcome occurs.

        For the present work, we followed the above methodology to calculate linear weights using all regular season MLB data from 2022 through 2024.
      
      \subsubsection{Hit Outcome Model}
      \label{sec:hit-outcome-model}

        A hit outcome model is a regression model which predicts the outcome of a fair ball based on batted ball tracking data. For example, most hard-hit line drives result in hits, but a batter might hit a hard line drive that results in an out because it is unluckily directed right where a fielder is positioned.

        We use the gradient boosting hit outcome model from \textcite{powers_pitch_2023}. For this model, the response variable is the change in base-out run expectancy resulting from the batted ball, and the features are batted ball exit speed, vertical/launch angle and horizontal/spray angle. It is trained on all MLB regular season batted balls from 2022 through 2024. The hyperparameters are reported in Table \ref{tab:xgboost-hyperparameters}. We call the predictions from this model {\it expected linear weight} (xLW).

      \subsubsection{Pitch Outcome Model}
      \label{sec:pitch-outcome-model}

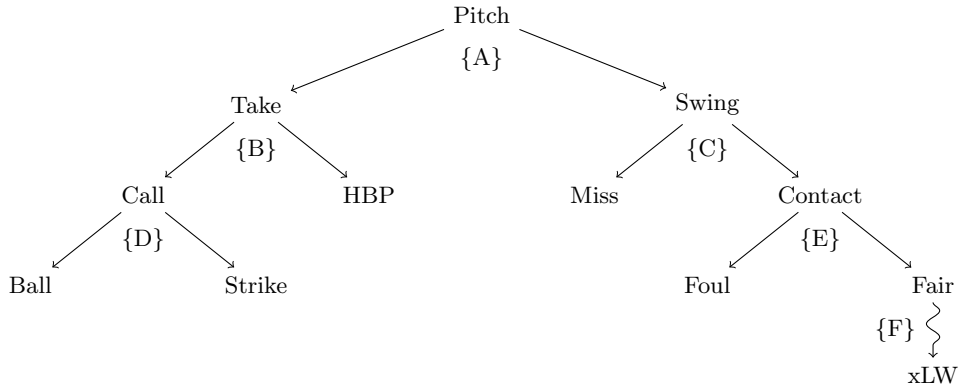
\begin{figure}[H]
          \centering
          \begin{tikzpicture}
            \node (pitch) at (0, 0) {\small Pitch};
            \node (A) at (0, -0.6) {\small \{A\}};
            \node (take) at (-3, -1.2) {\small Take};
            \node (B) at (-3, -1.8) {\small \{B\}};
            \node (swing) at (3, -1.2) {\small Swing};
            \node (C) at (3, -1.8) {\small \{C\}};
            \node (call) at (-4.5, -2.4) {\small Call};
            \node (D) at (-4.5, -3.0) {\small \{D\}};
            \node (hbp) at (-1.5, -2.4) {\small HBP};
            \node (miss) at (1.5, -2.4) {\small Miss};
            \node (contact) at (4.5, -2.4) {\small Contact};
            \node (E) at (4.5, -3.0) {\small \{E\}};
            \node (ball) at (-6, -3.6) {\small Ball};
            \node (strike) at (-3, -3.6) {\small Strike};
            \node (foul) at (3, -3.6) {\small Foul};
            \node (fair) at (6, -3.6) {\small Fair};
            \node (R) at (5.5, -4.2) {\small \{F\}};
            \node (xlw) at (6, -4.8) {\small xLW};
            \draw[->] (pitch) -- (take);
            \draw[->] (pitch) -- (swing);
            \draw[->] (take) -- (call);
            \draw[->] (take) -- (hbp);
            \draw[->] (swing) -- (miss);
            \draw[->] (swing) -- (contact);
            \draw[->] (call) -- (ball);
            \draw[->] (call) -- (strike);
            \draw[->] (contact) -- (foul);
            \draw[->] (contact) -- (fair);
            \draw[->, decorate, decoration = snake] (fair) -- (xlw);
          \end{tikzpicture}
          \caption{\it A pitch outcome tree diagram for the components of the pitch outcome model. Each split in the tree is a binary split with a corresponding binary gradient boosting model, except for the final split, corresponding to component model \{F\}, for which the outcome is continuous.}
          \label{fig:pitch-outcome-model}
        \end{figure}

        We utilize the pitch outcome model from \textcite{powers_pitch_2023}. This is a collection of six component models, as illustrated in Figure \ref{fig:pitch-outcome-model}. Each component is a gradient boosting model with five contextual features (balls, strikes, batter side, strike zone top, and strike zone bottom) and nine features from the trajectory of the pitch (2D location at plate, 3D velociy vector at plate, 3D acceleration vector at plate, 1D extension at release). Five of the component models predict probabilities of binary outcomes, and one component model predicts the expectation of a continuous outcome.

        Table \ref{tab:xgboost-hyperparameters} provides details of the pitch outcome component models. We trained the pitch outcome model using the \texttt{R} package \texttt{predpitchscore} (\cite{powers_predpitchscore_2025}), which relies on the eXtreme Gradient Boosting (XGBoost) algorithm (\cite{chen_xgboost_2016}). Models were trained on MLB regular season pitches from 2022 through 2024 with 5-fold cross-validation for hyperparameter tuning.
  
        \begin{table}
          \hspace{-6mm}
          \begin{tabular}{cl|rrrr}
                  &                                             & \multicolumn{4}{c}{XGBoost Hyperparameters}\\
            Model & Description                                 & \texttt{nrounds}  & \texttt{eta}        & \texttt{max\_depth} & \texttt{min\_child\_weight}\\
            \hline        
                  & Hit outcome model                           & 1000              & 0.05                & 9                   & 100\\
            \hline        
                  & {\it Pitch outcome model}\\
            \{A\} & Probability of swing                        & 1500              & 0.05                & 9                   & 10\\
            \{B\} & Probability of HBP given no swing  &  400              & 0.05                & 6                   & 10\\
            \{C\} & Probability of contact given swing          & 1000              & 0.01                & 6                   & 100\\
            \{D\} & Probability of strike given called pitch    & 2000              & 0.01                & 9                   & 10\\
            \{E\} & Probability of fair ball given contact      & 1500              & 0.01                & 9                   & 100\\
            \{F\} & Expected xLW given fair ball                & 1000              & 0.01                & 6                   & 100\\
            \hline
                & Hyperparameter search space                   & (0, 2000]         & \{0.01, 0.05, 0.3\} & \{3, 6, 9\}         & \{10, 30, 100\}
          \end{tabular}
          \caption{\it Descriptions and hyperparameters for hit outcome model and pitch outcome model components. Gradient boosting hyperparameters were tuned using 5-fold cross-validation, except for \texttt{gamma} (0), \texttt{subsample} (0.65), and \texttt{colsample\_by\_tree} (0.7), which were fixed in advance.}
          \label{tab:xgboost-hyperparameters}
        \end{table}
        
  \section{Methods}
  \label{sec:methods}

    \subsection{Intention Model}
    \label{sec:methods-intention}

    Acknowledging that swing timing biases the measurement of bat speed and swing length (see Figure \ref{fig:counterintuitive}), we first filter our dataset to consider only swings that result in squared-up contact against pitchers' primary fastballs. In effect, this is a form of ``matching'' to mitigate the confounding bias of timing and pitch recognition on the observed bat speed and swing length measurements. We assume that this filtered dataset of 32,094 high-quality contact swings against primary fastballs captures the batter's \textit{intention}. Thus, we consider models fit to this filtered dataset as a way of capturing variation in the batter's \textit{intended} swing length and bat speed.

\begin{figure}[H]
      \centering
      \includegraphics[width = 0.8\textwidth]{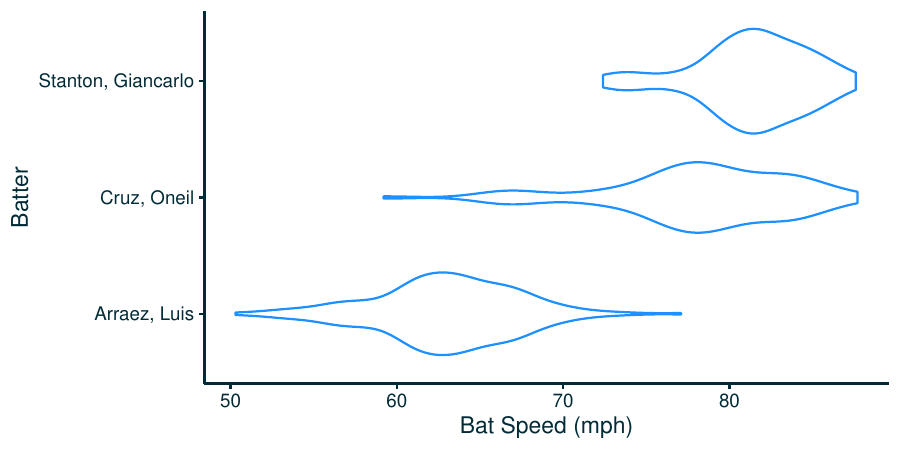}
      \caption{\it Distribution of bat speed for three different example batters, considering only swings which resulted in squared-up contact against pitchers' primary fastballs.}
      \label{fig:ex-bat-speed}
    \end{figure}

    We hypothesize that two covariates explain real (non-artifactual) differences in swing mechanics: ball-strike count and pitch location. To the extent that a batter's swing tracking metrics co-vary with count, we describe this as their \textit{approach}. To the extent that a batter's swing tracking metrics co-vary with pitch location, we describe this as swing \textit{adaptation}.  Additionally, based on visualizations of batter swing length and bat speed distributions (such as examples in Figure \ref{fig:ex-bat-speed}) we are also interested in capturing variation in the shape of the intended distributions for batters. 

    To this end, we fit the following skew-normal (SK) multilevel model:
    \begin{align}
    \label{eqn:intention-swing-length}
    \begin{split}
        ( \mbox{swing length} )_i &\sim \mbox{SK}(\mu_i, \sigma, \alpha_i) \\
        \mu_i &= \mu_0 + \gamma_{p_i} + \gamma_{b_i}
        + \beta^B \cdot (\mbox{balls})_i
          + (\beta^S + \gamma^S_{b_i}) \cdot (\mbox{strikes})_i\\
        & + (\beta^X + \gamma^X_{b_i}) \cdot (\mbox{pitch loc x})_i
          + (\beta^Z + \gamma^Z_{b_i}) \cdot (\mbox{pitch loc z})_i \\
          \alpha_i &= \alpha_0 + \nu_{b_i}
    \end{split}
    \end{align}
    where $b_i$ denotes the batter on swing $i$ against pitcher $p_i$. We use the same specification for both swing length and bat speed intention models. 
    
    In detail, the $\gamma$ parameters are random intercepts and random slopes describing the mean of the swing distribution such that
    \begin{align}
        \begin{split}
            \gamma_{p_i} &\sim \mathcal{N}(0, \sigma^2_p) \\
            \boldsymbol{\gamma}_{b_i} &\sim \mathcal{N}(\boldsymbol{0}, \Sigma)
        \end{split}
    \end{align}
    where $\boldsymbol{\gamma}_{b_i}$ is the vector of batter random intercept and slopes $(\gamma_{b_i}, \gamma^S_{b_i}, \gamma^X_{b_i}, \gamma^Z_{b_i})$ which follow a multivariate normal distribution centered around the zero vector $\boldsymbol{0}$ with covariance matrix $\Sigma$. This $\Sigma$ contains the relevant random effect variances for the batter-level terms, $(\sigma_b, \sigma_b^S, \sigma_b^X, \sigma_b^Z)$, as well as their respective correlations. This allows us to quantify variation between batters and pitchers at the intercept level, as well as capturing variation in batter approach and swing adaptation.
    
    The $\nu_{b_i}$ parameters are random intercepts enabling us to capture variation between batters in the shape $\alpha$ of the swing length and bat speed distributions, where
    \begin{equation}
        \nu_{b_i} \sim \mathcal{N}(0, \tau^2_b).
    \end{equation}

    These models tell us: What are the bat speed and swing length (by batter, count and pitch location) when the timing is good? We interpret the prediction from this model on each pitch as the intended bat speed and swing length.
    
    We fit both bat speed and swing length models in a Bayesian framework via the \texttt{R} package \texttt{brms}  (\cite{burkner_bayesian_2021}), which provides an interface for Bayesian modeling with Stan (\cite{carpenter_stan_2017}). We use weakly informative priors for the parameters, with vague half-$t_3$ priors (i.e., a Student's $t$ distribution centered at zero with 3 degrees of freedom but truncated to positive values) for the standard deviation parameters (\cite{gelman_prior_2006}) and the LKJ prior for the correlations between batter-level effects (\cite{lewandowski_generating_2009}). Our Bayesian approach naturally provides uncertainty quantification for the model parameters via their posterior distributions, which are estimated using MCMC sampling. For model fitting, we use four parallel chains with 6,000 (3,000 burn-in) and 4,000 (2,000 burn-in) samples for the swing length and bat speed models, respectively. We observe evidence of convergence of the MCMC algorithm based on trace plots and $\hat{R}$ statistics close to 1 (\cite{gelman_inference_1992, brooks_general_1998}), along with no indication of problematic effective sample sizes (\cite{gelman_bayesian_2020}).
    
    For comparison, we also fit simpler Gaussian versions of the swing length and bat speed models. These models follow a specification similar to that of the SK models in Equation (\ref{eqn:intention-swing-length}), but with the shape parameter removed (i.e., the Gaussian distribution is a SK with shape 0). We use an 80/20 train/test split to compare the out-of-sample performance of the SK and Gaussian fits against each other. We compute the expected log pointwise predictive density (ELPD) on the 20\% test data, which involves computing the log-likelihood of each test observation for each posterior sample. In addition to providing us with standard errors about the difference in model performance, we choose ELPD as our evaluation metric because of our interest in comparing the skewness of the SK posterior predictive distributions relative to the symmetric Gaussian.

    \subsection{Causal Model}
    \label{sec:methods-causal}

      The intention model from Section \ref{sec:methods-intention} sets us up to estimate a causal model for the effect of bat speed and swing length on swing outcomes via instrumental variables regression (\cite{bollen_instrumental_2012}). Relying on baseball domain knowledge, we hypothesize that the relationship between swing tracking metrics and outcomes is confounded by a latent, unobserved variable: whether the batter correctly identifies the pitch type early in its trajectory (by lucky guess or by other means) or needs to make a mid-swing adjustment. When the batter fails to correctly identify the pitch type early, he is less likely to make good contact, and he also adjusts his swing, reducing the bat speed and affecting the swing length. Critically, this confounder can vary significantly from pitch to pitch as batters frequently limit their focus to one or two pitch types because it is too difficult to prepare for any pitch type (\cite{gray_markov_2002}).

      A good instrumental variable is one which has an effect on the treatment (in this case, swing tracking metrics) and is known to be uncorrelated with the confounder (\cite{bollen_instrumental_2012}). As the instrument varies, we observe the effect of the treatment with the effect of the confounder stripped away. Relying on the framework from Section \ref{sec:methods-intention}, we choose the batter's count-intended bat speed and swing length as our instrumental variables. Count-intended swing length is the player-specific contribution of count to swing length, as estimated by (\ref{eqn:intention-swing-length}), and count-intended bat speed is defined analogously.

      Count-intended swing metrics affect observed swing metrics as modeled in Section \ref{sec:methods-intention}, and we expect that they are approximately uncorrelated with early pitch recognition. The count is known before the pitch and does not depend on the pitch itself. Precise independence may not be a valid assumption, for example, if batters who tend to struggle more with early pitch recognition in two-strike counts tend also to modify their two-strike swing intentions a certain way. But this choice of instrumental variables mitigates any confounding effect of early pitch recognition from pitch to pitch.

      We consider the effect of bat speed and swing length on three components of the pitch outcome model (see Figure \ref{fig:pitch-outcome-model}): \{C\} probability of contact conditioned on swing; \{E\} probability of fair ball conditioned on contact; and \{F\} expected xLW conditioned on fair ball (xLW is explained in Section \ref{sec:hit-outcome-model}). For each of these components, we estimate a logistic or linear regression of the outcome on the batter's count-intended bat speed and swing length, with an offset for the prediction from the pitch outcome model described in Section \ref{sec:pitch-outcome-model}. This offset controls for the difficulty of each pitch when estimating the effects of count-based bat speed and swing length adjustments.

      Using $i \in \{1, ..., n\}$ to index swings, the random variables $Y_i^{\mbox{\scriptsize con}} \in \{0, 1\}$, $Y_i^{\mbox{\scriptsize fair}} \in \{0, 1\}$ and $Y_i^{\mbox{\scriptsize xLW}} \in \mathbb{R}_+$ represent whether swing $i$ results in contact; whether swing $i$ results in a fair ball; and the xLW of the batter ball on swing $i$ (undefined if $Y_i^{\mbox{\scriptsize fair}} \in \{0, 1\} = 0$), respectively. The corresponding predictions from the pitch outcome model are denoted by $\hat p_i^{\mbox{\scriptsize con}} \in (0, 1)$,  $\hat p_i^{\mbox{\scriptsize fair}} \in (0, 1)$ and $\hat y_i^{\mbox{\scriptsize xLW}} \in \mathbb{R}$. We use $b_i$ to denote the batter on swing $i$, and $\hat\gamma_{b_i}^{\mbox{\scriptsize BS}}$ and $\hat\gamma_{b_i}^{\mbox{\scriptsize SL}}$ are the estimated batter random slopes for strikes in the bat speed intention model and the swing length intention model (\ref{eqn:intention-swing-length}), respectively.

      \begin{align}
        \label{eqn:causal-contact}
        \log \left(
          \frac{\mathbb{P}(Y_i^{\mbox{\scriptsize con}} = 1)}{1 - \mathbb{P}(Y_i^{\mbox{\scriptsize con}} = 1)}
        \right) &= \log \left(
          \frac{\hat p_i^{\mbox{\scriptsize con}}}{1 - \hat p_i^{\mbox{\scriptsize con}}}
        \right) +
          \alpha^{\mbox{\scriptsize con}} +
          \beta_{\mbox{\scriptsize BS}}^{\mbox{\scriptsize con}} \cdot
            \hat\gamma_{b_i}^{\mbox{\scriptsize BS}} +
          \beta_{\mbox{\scriptsize SL}}^{\mbox{\scriptsize con}} \cdot
            \hat\gamma_{b_i}^{\mbox{\scriptsize SL}},\\[10pt]
        \label{eqn:causal-fair}
        \log \left(
          \frac{
            \mathbb{P}(Y_i^{\mbox{\scriptsize fair}} = 1 \mid Y_i^{\mbox{\scriptsize con}} = 1)
          }{
            1 - \mathbb{P}(Y_i^{\mbox{\scriptsize fair}} = 1 \mid Y_i^{\mbox{\scriptsize con}} = 1)
          }
        \right) &= \log \left(
          \frac{\hat p_i^{\mbox{\scriptsize fair}}}{1 - \hat p_i^{\mbox{\scriptsize fair}}}
        \right) +
          \alpha^{\mbox{\scriptsize fair}} +
          \beta_{\mbox{\scriptsize BS}}^{\mbox{\scriptsize fair}} \cdot
            \hat\gamma_{b_i}^{\mbox{\scriptsize BS}} +
          \beta_{\mbox{\scriptsize SL}}^{\mbox{\scriptsize fair}} \cdot
            \hat\gamma_{b_i}^{\mbox{\scriptsize SL}},\\[10pt]
        \label{eqn:causal-hit}
        Y_i^{\mbox{\scriptsize xLW}} \mid \{Y_i^{\mbox{\scriptsize fair}} = 1\} &\sim
          \mbox{Normal}\left(
            \hat y_i^{\mbox{\scriptsize xLW}} +
              \alpha^{\mbox{\scriptsize xLW}} +
              \beta_{\mbox{\scriptsize BS}}^{\mbox{\scriptsize xLW}} \cdot
                \hat\gamma_{b_i}^{\mbox{\scriptsize BS}} +
              \beta_{\mbox{\scriptsize SL}}^{\mbox{\scriptsize xLW}} \cdot
                \hat\gamma_{b_i}^{\mbox{\scriptsize SL}},
            \sigma^2
          \right).
      \end{align}

      In each of these models, we interpret $\alpha^\cdot$ as an intercept to account for the fact that the mean outcome may not be the same in this dataset as in the dataset on which the pitch outcome model was trained. We interpret $\beta_{\mbox{\scriptsize BS}}^{\cdot}$ and $\beta_{\mbox{\scriptsize SL}}^{\cdot}$ as the effects of bat speed and swing length adjusments on the outcomes. The parameter $\sigma^2$ in the xLW model is a nuisance parameter.

      The Gaussian distribution assumed by (\ref{eqn:causal-hit}) is likely not true. A substantial portion of batted balls are routine outs (xLW $\approx -0.27$, the minimum) while some batted balls are likely home runs (xLW $> 1$). The unconditional distribution of xLW is right-skewed with skewness 1.55. This violation of the Gaussian assumption does not cause problems for the point estimates of the regression coefficients in (\ref{eqn:causal-hit}), but it does cause our methodology to under-estimate their standard errors.

    \subsection{Run Value Model}
    \label{sec:methods-value}

      The models from Section \ref{sec:methods-causal} estimate the effect of swing adjustments on contact and power, but they don't help us make value judgments when these effects move in opposite directions, e.g. sacrificing power for increased contact. In baseball, the units of value are runs, which translate directly to wins (\cite{james_bill_1983}). The standard metric for estimating the run value of batting performance is linear weight, which assigns a run value to each type of batter outcome (strikeout, walk, home run, etc.) according to its average change in base-out run expectancy (\cite{thorn_hidden_1984}). In this section, we describe a method for converting the effects on contact and power into effects on linear weight.

      We model the progression of a plate appearance as a Markov reward process whose nonterminal states are all possible ball-strike counts $\mathcal{S}_N \equiv \{0, 1, 2, 3\} \times \{0, 1, 2\}$ and whose terminal states are $\mathcal{S}_T \equiv$ \{Strikeout, Walk, Hit by Pitch, Fair Ball\}. From Section \ref{sec:methods-causal}, we have the means by which to estimate the probability of each possible outcome of a pitch given its trajectory and the batter's approach, assuming an otherwise average batter. These probabilities induce transition probabilities between the states $\mathcal{S} \equiv \mathcal{S}_N \cup \mathcal{S}_T$ of the Markov reward process because the pitch outcome (HBP, Called Ball, Called Strike, Swinging Strike, Foul Ball or Fair Ball) probability is the product of conditional probabilities from models \{A\}--\{E\}, and the state transition is a deterministic function of the pitch outcome. We use $\hat q(s' | s, \vec x, \vec \gamma)$ to denote the estimated transition probability from $s \in \mathcal{S}$ to $s' \in \mathcal{S}$ induced by the estimated pitch outcome probabilities given pitch trajectory features $\vec x \in \mathbb{R}^9$ and batter approach $\vec\gamma \equiv (\gamma^{\mbox{\scriptsize BS}}, \gamma^{\mbox{\scriptsize SL}}) \in \mathbb{R}^2$.

      Our goal is to estimate the expected run value of a plate appearance given any batter approach $\gamma$. We assume that the trajectory of each pitch in the plate appearance is drawn from the leaguewide empirical distribution of pitch trajectories in the 2024 MLB regular season, conditioned on the ball-strike count. Indexing these pitches by $i = 1, ..., n$, we use $s_i \in \mathcal{S}_N$ to denote the ball-strike count of pitch $i$, and we use $\vec x_i \in \mathbb{R}^9$ to denote the trajectory features of pitch $i$. Under the foregoing assumption, the transition probability function given batter approach $\vec\gamma \in \mathbb{R}^2$ is
      \begin{equation*}
        \tilde q(s' | s, \vec\gamma) = \frac{
          \sum_{i = 1}^n \mathbb{I}\{s_i = s\} \hat q(s' | s, \vec x_i, \vec\gamma)
        }{
          \sum_{i = 1}^n \mathbb{I}\{s_i = s\}
        } \mbox{ for }s \in \mathcal{S}_N,\, s' \in \mathcal{S}.
      \end{equation*}

      For three terminal states, we used fixed linear weights calculated using 2022-2024 MLB data (as detailed in Section \ref{sec:linear-weights}): LW(Strikeout) = --0.27, LW(Walk) = +0.33, and LW(Hit by Pitch) = +0.36. The linear weight of the final terminal state, Fair Ball, depends on the count and the batter approach through the pitch outcome model predictions. The earlier assumption that the trajectory of each pitch is drawn from the leaguewide distribution of trajectories, conditioned on count, implies that
      \begin{equation*}
        \mbox{LW}(\mbox{Fair Ball} \mid s, \vec\gamma) = \frac{
          \sum_{i = 1}^n \mathbb{I}\{s_i = s\} \hat E[Y_i^{\mbox{\scriptsize xLW}} | s_i, \vec x_i,\, \vec\gamma]
        }{
          \sum_{i = 1}^n \mathbb{I}\{s_i = s\}
        }.
      \end{equation*}

      Finally, we define the reward function of the Markov reward process as
      \begin{equation*}
          r(s' | s, \vec\gamma) = \begin{cases}
            \mbox{LW}(s) & \mbox{ if } s' \in \{\mbox{Strikeout, Walk, Hit By Pitch}\}\\
            \mbox{LW}(\mbox{Fair Ball} \mid s, \vec\gamma) & \mbox{ if } s' = \mbox{Fair Ball}
          \end{cases}.
      \end{equation*}
      We calculate the value of each state by iteratively solving the Bellman equation (\cite{bellman_dynamic_1957}); and the corresponding value of the initial state $s = (0, 0)$ gives us the estimated run value of the approach $\vec\gamma$.

  \section{Results}
  \label{sec:results}

    \subsection{Intention Model}
    \label{sec:results-intention}

      The out-of-sample ELPD results are shown in Table \ref{tab:intention-model-elpd} for both swing length and bat speed models. For both settings, the SK model displayed an ELPD of around three standard errors better than the Gaussian model. The remaining results in this manuscript are based on the SK intention models.

      \begin{table}[H]
        \centering
        \begin{tabular}{l|r|r|r|}
                & $\Delta$ELPD  & SE    & \# SEs \\
  \hline
  Swing length  & 34.13         & 11.72 & 2.91 \\ 
  Bat speed     & 87.60         & 27.88 & 3.14
\end{tabular}
        \caption{\it Model comparison via expected log predictive density (ELPD). Shown are the difference in ELPD between the skew-normal (SK) and Gaussian models ($\Delta$ELPD), the standard error of ELPD difference (SE), and the number of standards errors by which the SK model outperforms the Gaussian model.}
        \label{tab:intention-model-elpd}
      \end{table}

\begin{figure}
        \centering
        \includegraphics[width = 0.6\textwidth]{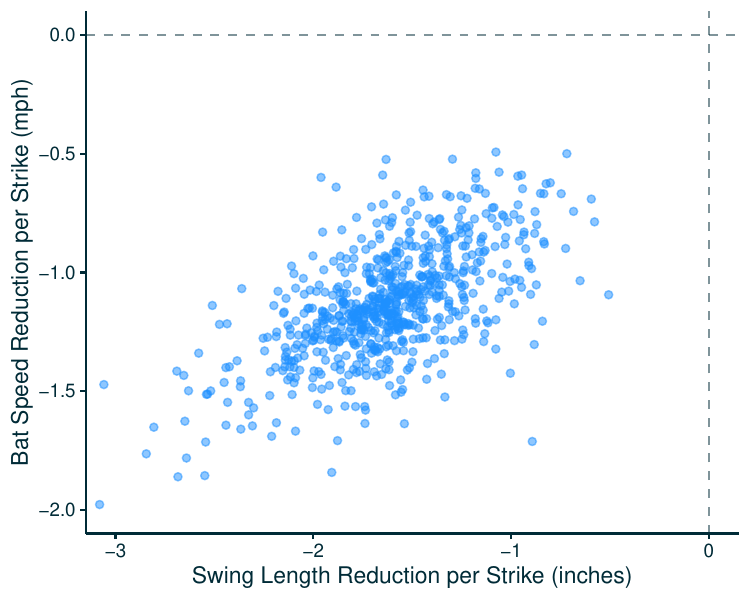}
        \caption{\it Estimated batter random slopes for strikes in the intention model. Each point is a batter $b$, with the x-value representing $\hat\beta^S + \hat\gamma^S_b$ from the model (\ref{eqn:intention-swing-length}) and the y-value representing the same quantity from the analogous model for intended bat speed. This quantity is interpretable as the expected change in swing length (or bat speed) corresponding to a one-strike increase in the ball-strike count.}
        \label{fig:approach}
      \end{figure}

       Table \ref{tab:fixed-effects} summarizes the fixed effects of the bat speed and swing length intention models. We observe intuitive relationships for both models with regard to the relationships between pitch-level context and each observed swing measurement. As the number of balls increase ($\beta^B$), on average we observe faster and longer swings. The opposite behavior is observed as the number of strikes increases ($\beta^S$), with slower and shorter swings on average. Figure \ref{fig:approach} displays the joint distribution of the posterior mean for the batter random slopes for strikes ($\hat{B}^S + \hat{\gamma}_b^S$) in both intention models. Regardless of the individual batter effect, we expect a decrease in bat speed and swing length per strike. A similar relationship is observed for pitches with higher vertical locations ($\beta^Z$). The two models diverge with regard to the horizontal location of pitches ($\beta^X$), as pitches that are thrown further away from the batter result in longer but slower swings on average. With regard to the shape parameter of the SK distribution, the intercepts ($\alpha_0$) for both bat speed and swing length are negative, which indicate left-skewed distributions.

      \begin{table}[H]
        \centering
        \begin{tabular}{l|rrr|rrr|}
    & \multicolumn{3}{c|}{Bat Speed}          & \multicolumn{3}{c|}{Swing Length}  \\
  Parameter & Mean  & Lower  & Upper & Mean  & Lower  & Upper \\
  \hline
  $\mu_0$ & 76.43 & 76.06 & 76.80 & 8.25 & 8.20 & 8.29 \\
  $\beta^B$ & 0.55 & 0.51 & 0.59 & 0.05 & 0.04 & 0.05 \\
  $\beta^S$ & --1.13 & --1.20 & --1.06 & --0.14 & --0.14 & --0.13 \\
  $\beta^X$ & --0.52 & --0.66 & --0.39 & 0.16 & 0.14 & 0.17 \\
  $\beta^Z$ & --1.87 & --2.00 & --1.75 & --0.46 & --0.47 & --0.44 \\
  $\alpha_0$ & --1.85 & --1.96 & --1.75 & --1.46 & --1.56 & --1.36 \\
\end{tabular}
        \caption{\it Posterior mean and 95\% credible interval lower/upper bounds for the fixed effects in the bat speed and swing length intention models.}
        \label{tab:fixed-effects}
      \end{table}

      Next, Table \ref{tab:intention-variances} summarizes the posterior distributions of the random effect variance terms. For both bat speed and swing length intention models, we observe that the highest variation is attributed to differences between batters ($\sigma_b$). Interestingly, the variation between batters in the shape random effects ($\tau_b$) is the second highest for swing length but is smaller than the variation in the pitch location random slopes ($\sigma_b^X$ and $\sigma_b^Z$) for the bat speed model. The estimated terms demonstrate evidence of variability between batters at the intercept and slope level.
       
      \begin{table}[H]
        \centering
        \begin{tabular}{c|rrr|rrr|}
    & \multicolumn{3}{c|}{Bat Speed}          & \multicolumn{3}{c|}{Swing Length}  \\
  Parameter & Mean  & Lower  & Upper & Mean  & Lower  & Upper \\
  \hline
  $\sigma_p$ & 0.47 & 0.41 & 0.53 & 0.12 & 0.11 & 0.13 \\
  $\sigma_b$ & 3.35 & 3.04 & 3.67 & 0.43 & 0.40 & 0.47 \\
  $\sigma_b^S$ & 0.42 & 0.35 & 0.50 & 0.05 & 0.05 & 0.06 \\
  $\sigma_b^X$ & 1.04 & 0.90 & 1.18 & 0.11 & 0.10 & 0.13 \\
  $\sigma_b^Z$ & 1.05 & 0.94 & 1.17 & 0.13 & 0.11 & 0.14 \\
  $\tau_b$ & 0.67 & 0.52 & 0.85 & 0.34 & 0.08 & 0.57 \\
\end{tabular}
        \caption{\it Posterior mean and 95\% credible interval lower/upper bounds for the standard deviation of the random effect terms in the bat speed and swing length intention models.}
        \label{tab:intention-variances}
      \end{table}

\begin{figure}[H]
        \centering
        \includegraphics[width = \textwidth]{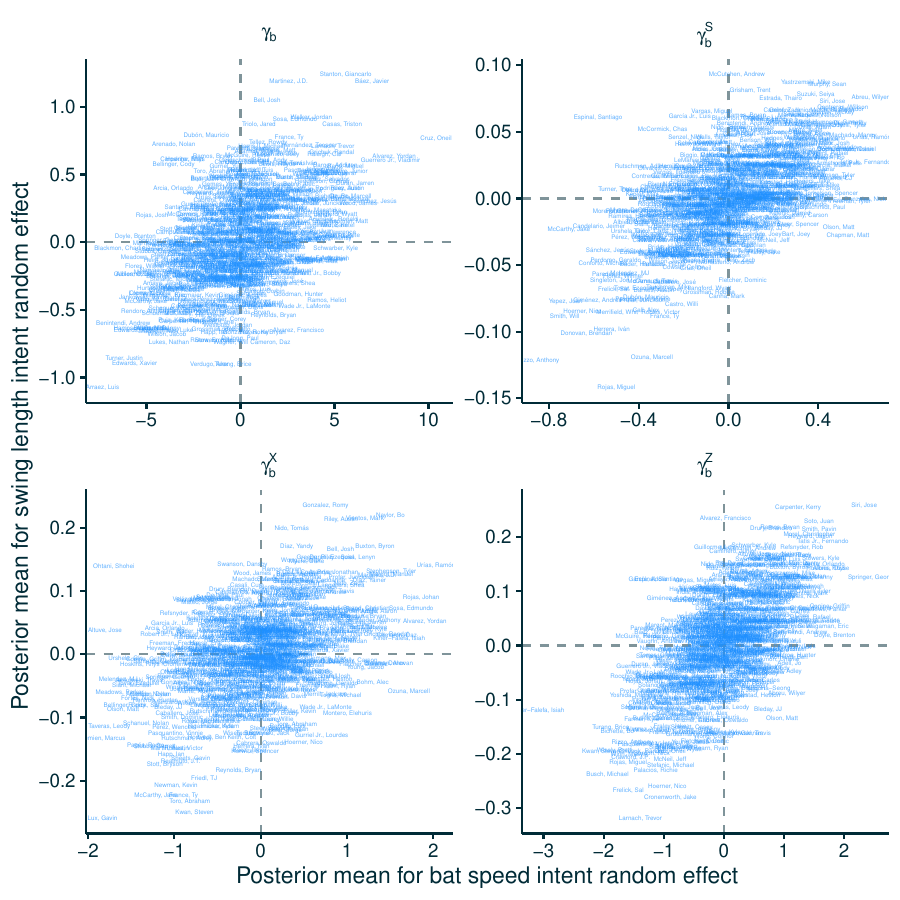}
        \caption{\it Joint distribution of the posterior means for the batter-level $\mu_i$ random effects in the SK intention models for both bat speed and swing length. Each batter is displayed by their name, and only batters with at least 25 squared-up swings are displayed.}
        \label{fig:intent-re}
      \end{figure}

      Figure \ref{fig:intent-re} provides an overview of the batter-level $\mu_i$ random effects in the SK intention models. The variation we observe between batters aligns with intuition, as batters known for taking longer/faster swings (e.g., Oneil Cruz, Giancarlo Stanton, Javier Baez) display higher random intercepts ($\gamma_b$) in comparison with batters known for shorter/slower swings (e.g., Luis Arraez). We observe positive correlations between the random slopes of bat speed and swing length, providing information on the variation between batters with respect to their approach ($\gamma_b^S$) and adaptation ($\gamma_b^X$, $\gamma_b^Z$). For example, Anthony Rizzo's approach appears to be more conservative, with the biggest decrease in bat speed and swing length for each additional strike relative to the average batter slope. We also observe batters that deviate from the consistent pattern between bat speed and swing length, such as Shohei Ohtani with regard to his horizontal pitch location random slope appearing to be higher than average for swing length despite a lower-than-average value for bat speed. 
      
\begin{figure}[H]
        \centering
        \includegraphics[width = 0.8\textwidth]{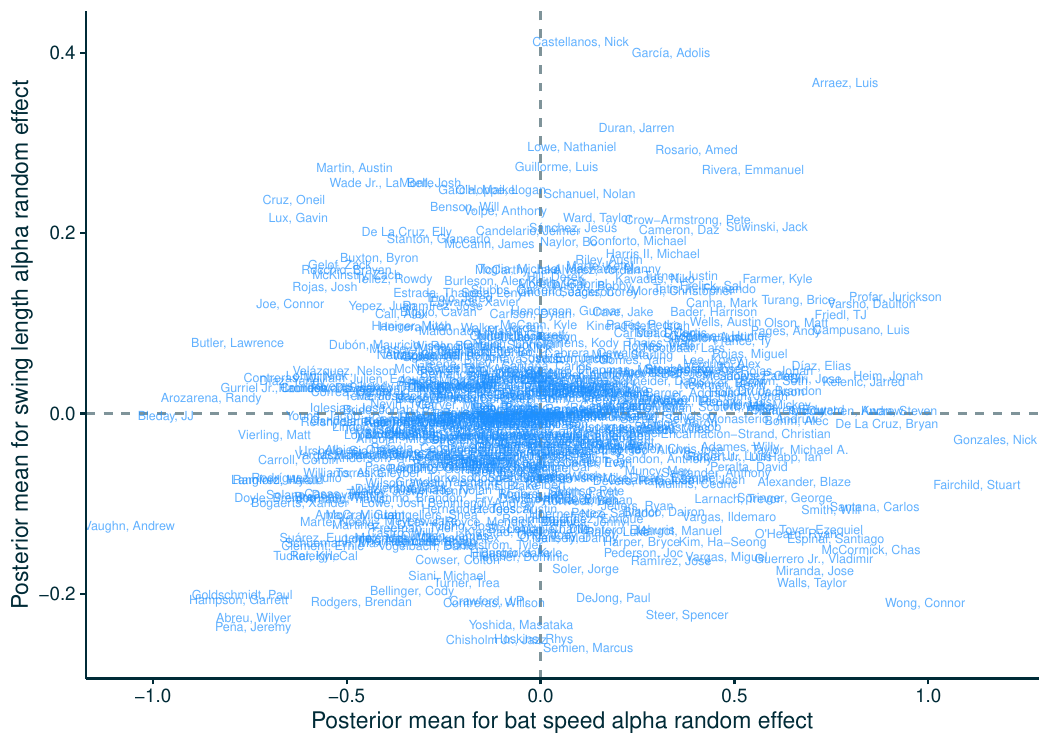}
        \caption{\it Joint distribution of the posterior means for the batter-level $\alpha_i$ random effects in the SK intention models for both bat speed and swing length. Each batter is displayed by their name, and only batters with at least 25 squared-up swings are displayed.}
        \label{fig:alpha-re}
      \end{figure}

       In contrast, Figure \ref{fig:alpha-re} displays the joint distribution of the posterior means for the batter random intercepts in modeling the shape parameter $\alpha$ of the SK distribution. Unlike the mean-level random effects, we do not observe correlation between the bat speed and swing length shape-level random intercepts. The variation captured by these random intercepts provides some understanding of the differences in the intended swing distribution for players. Luis Arraez displays above-average random intercepts for both bat speed and swing length, indicating that his intention distributions are less skewed left than those of other batters. 
                
    \subsection{Causal Model}
    \label{sec:results-causal}

      The previous section demonstrates that there are differences between batters in how they modulate their swing length and bat speed as the number of strikes in the count increases. This section addresses the consequences of those differences between batters. As described in Section \ref{sec:methods-causal}, we estimate three separate regression models with instrumental variables for contact probability given swing (\ref{eqn:causal-contact}), fair ball probability given contact (\ref{eqn:causal-fair}) and expected xLW given fair ball (\ref{eqn:causal-hit}). The estimated effects of bat speed approach and swing length approach from these models are reported in Table \ref{tab:causal-model}.

      \begin{table}[H]
        \centering
        \begin{tabular}{l|r|r|r|}
 & Contact Model (\ref{eqn:causal-contact}) & Fair/Foul Model (\ref{eqn:causal-fair}) & xLW Model (\ref{eqn:causal-hit}) \\
  \hline
Bat Speed Approach (mph) & $-0.176 \pm0.016$ & $-0.074 \pm0.014$ & $0.022 \pm0.003$ \\ 
  Swing Length Approach (inches) & $-0.050 \pm0.010$ & $0.003 \pm0.009$ & $0.000 \pm0.002$ \\ 
\end{tabular}

        \caption{\it Estimated regression coefficients and corresponding standard errors from the causal models (\ref{eqn:causal-contact})--(\ref{eqn:causal-hit}). Swing Length Approach is defined as the batter's change in swing length per strike added to the count, as estimated by (\ref{eqn:intention-swing-length}). Bat Speed Approach is defined as the batter's change in bat speed per strike added to the count, as estimated by the analagous model for intended bat speed.}
        \label{tab:causal-model}
      \end{table}

      From the contact model, the negative coefficient $-0.176$ for bat speed approach shows that batters who reduce their bat speed more per strike (i.e. more negative bat speed approach) make contact more often in one-strike and two-strike counts. Similarly, the negative coefficient for swing length approach shows that batters who reduce their swing length more per strike make contact more often in these counts. These results match our intuition from baseball domain knowledge: More conservative swings are more likely to make contact, a more plausible conclusion than the counterintuitive results from the naive analysis shown in Figure \ref{fig:counterintuitive}. Bat speed approach and swing length approach have similar between-batter variance, so in comparing the coefficients we observed that bat speed approach has a much greater effect (more than 3x) on contact than does swing length approach. The fair ball probability model shows a similar effect direction for bat speed approach and no effect for swing length approach.

      From the xLW model (i.e. the ``power'' model), the positive coefficient $0.023$ for bat speed approach shows that batters who reduce their bat speed more per strike (i.e. more negative bat speed approach) exhibit less power in one-strike and two-strike counts when they do make contact. This demonstrates evidence for the intuitive tradeoff involved when taking more conservative swings: greater contact but less power. Swing length approach does not seem to have an effect on power.

      The magnitudes of the coefficients in Table \ref{tab:causal-model} are difficult to interpret because two of the models are logistic regressions, so the coefficients represent additive effects in the log-odds scale. For example, a pitch with almost 0\% contact probability will still have almost 0\% contact probability, regardless of approach. The effect of approach on contact is maximized for pitches near 50\% contact probability. Figure \ref{fig:approach} conveys the average effect size across all 2-strike pitches. We observe that the approach effects for almost all batters (relative to an average approach) are within $\pm4\%$ contact probability and $\pm.04$ xLW.

      \begin{figure}[H]
        \centering
        \includegraphics[width = 0.49\textwidth]{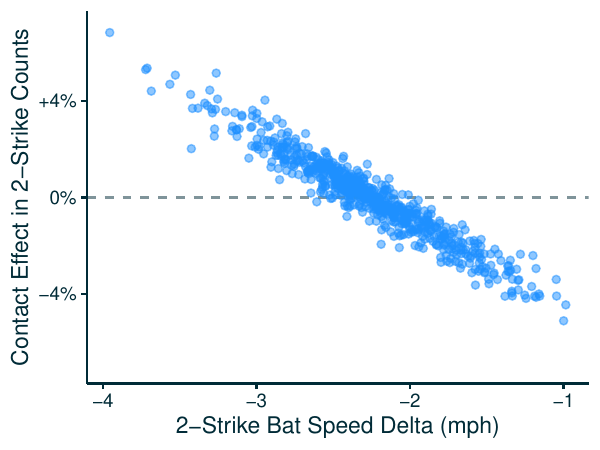}
        \includegraphics[width = 0.49\textwidth]{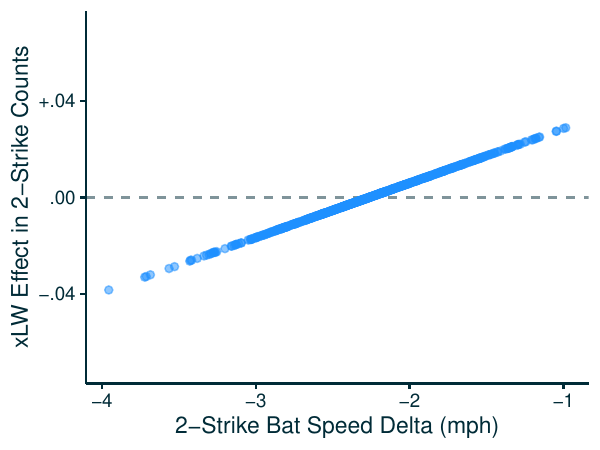}
        \caption{\it Estimated causal effect of bat speed on contact (left) and power (right). Each point is a batter; the x-axis shows how much the batter slows their swing in two-strike counts, relative to zero-strike counts; the y-axis shows the estimated combined effect of the batter's bat speed and swing length adjustments on contact rate (left) and xLW (right) in two-strike counts, from the models (\ref{eqn:causal-contact}) and (\ref{eqn:causal-hit}), respectively. xLW is expected linear weight, measured in runs per plate appearance (see Section \ref{sec:hit-outcome-model}).}
        \label{fig:results-causal}
      \end{figure}

    \subsection{Run Value Model}
    \label{sec:results-value}

\begin{figure}[H]
        \centering
        \includegraphics[width = 0.6\textwidth]{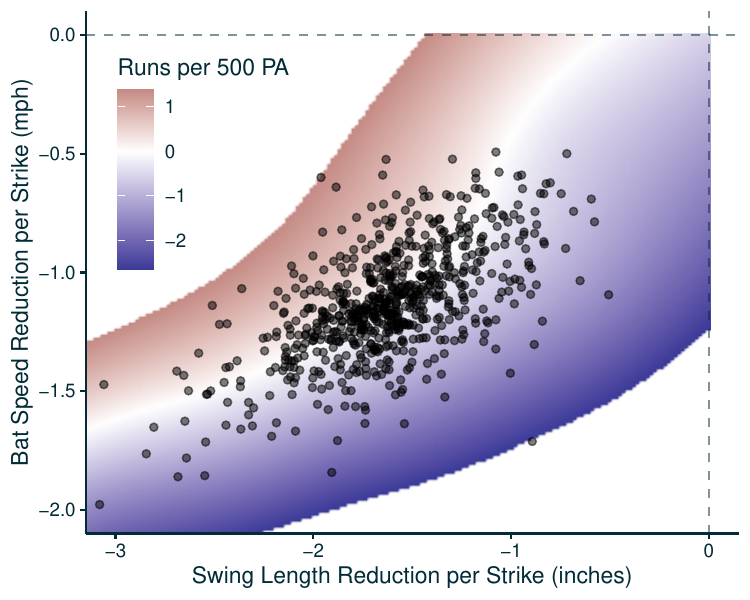}
        \caption{\it Estimated causal effect of batters' approaches, measured on the scale of runs per 500 plate appearances. This figure shows the same data as Figure \ref{fig:approach} except that the background gradient reports the estimated run value of each approach. If an approach is valued at z runs, the interpretation is that the average batter with this approach is expected to produce z runs more than if they adopted an average approach, as measured by linear weights.}
        \label{fig:approach-run-value}
      \end{figure}

      The previous section demonstrates that there is a contact/power tradeoff involved when batters modulate their swing length and bat speed as the number of strikes in the count increases. This section addresses whether the tradeoff is worthwhile. As described in Section \ref{sec:methods-value}, we estimate the expected run value of plate appearance outcomes for the average batter if they were to adopt different approaches. Figure \ref{fig:approach-run-value} plots batter approaches (as previously plotted in Figure \ref{fig:approach}) now with a color gradient in the background showing the estimated run value of each approach.

      The magnitude of the difference between the higest-value approaches and the lowest-value approaches is small, approximately 4 runs per 500 plate appearances. To put this in context, it amounts to a difference of roughly half a win per season (\cite{slowinski_converting_2010}). Baseball teams are chasing increasingly small competitive advantages, and the cost of half a win is roughly \$4 million in the free agent market (\cite{clemens_what_2021}). We observe that the highest-value approaches are those which cut down on swing length without cutting down on bat speed. From Table \ref{tab:causal-model}, we observe that this is a way to avoid the contact/power tradeoff because cutting down on swing length (all else equal) increases contact without decreasing power.

      Table \ref{tab:approach-ranked} reports the specific approaches and corresponding run values of the top 5 and bottom 5 batters, ranked according to the run value of their approaches {\it applied to an average batter}. Interestingly, the top three batters (Chapman, Olson and Canha) all spent the first five or more seasons of their MLB careers around the same time with the Oakland Athletics.

      \begin{table}
        \centering
        \begin{tabular}{rl|rr|r}
& & \multicolumn{2}{c|}{Approach} & Runs /\\
 & Batter & Bat Speed (mph) & Swing Length (in.) & 500 PA \\
  \hline
  1 & Matt Chapman & $-0.60$ & $-1.96$ & $1.39$ \\ 
    2 & Matt Olson & $-0.64$ & $-1.88$ & $1.21$ \\ 
    3 & Mark Canha & $-1.14$ & $-2.51$ & $1.13$ \\ 
    4 & Dominic Fletcher & $-1.07$ & $-2.36$ & $1.12$ \\ 
    5 & Joey Bart & $-0.83$ & $-1.95$ & $1.04$ \\ 
   &  &  &  &  \\ 
  688 & Jeimer Candelario & $-1.71$ & $-1.88$ & $-1.52$ \\ 
  689 & Chas McCormick & $-1.42$ & $-1.00$ & $-1.60$ \\ 
  690 & Trea Turner & $-1.64$ & $-1.54$ & $-1.62$ \\ 
  691 & Jake McCarthy & $-1.84$ & $-1.91$ & $-2.02$ \\ 
  692 & Santiago Espinal & $-1.71$ & $-0.89$ & $-2.65$ \\ 
\end{tabular}

        \caption{\it Top 5 and bottom 5 approaches, ranked by runs per 500 plate appearances. Bat Speed Approach and Swing Length Approach are the batter's change in bat speed and swing length, respectively, per strike added to the count. We estimate the effect that each approach would have on the average batter's performance, as measured on the run scale via linear weights.}
        \label{tab:approach-ranked}
      \end{table}

  \section{Discussion}
  \label{sec:discussion}

    In this work, we use newly available bat tracking data to examine the relationship between swing aggression and outcomes in baseball. We recognize that both bat speed and swing length are inherently outcomes of the swing, confounded by whether or not the batter identified the pitch. To address this, we first model the batter's bat speed and swing length intent via Bayesian hierarchical skew-normal models with evidence of more appropriate fits compared to simpler Gaussian models. Our results demonstrate the variation between batters in multiple aspects, including how they change their swing metrics as the number of strikes increases. We then condition on the batter's count-intended bat speed and swing length as instrumental variables to estimate the causal effect of these intended swing metrics on swing outcomes. The results are intuitive, indicating that batters who shorten and slow down their swing more per strike are expected to make contact more often in one-strike and two-strike counts relative to other batters. However, we observe a tradeoff as this type of conservative approach at the plate is associated with less power. We quantify this tradeoff in terms of run value, revealing on an interpretable scale that the highest-value approaches cut down on swing length without sacrificing bat speed. 

    \subsection{Other Sources of Swing Variation}
    \label{sec:results-other}

\begin{figure}
        \centering
        \includegraphics[width = 0.8\textwidth]{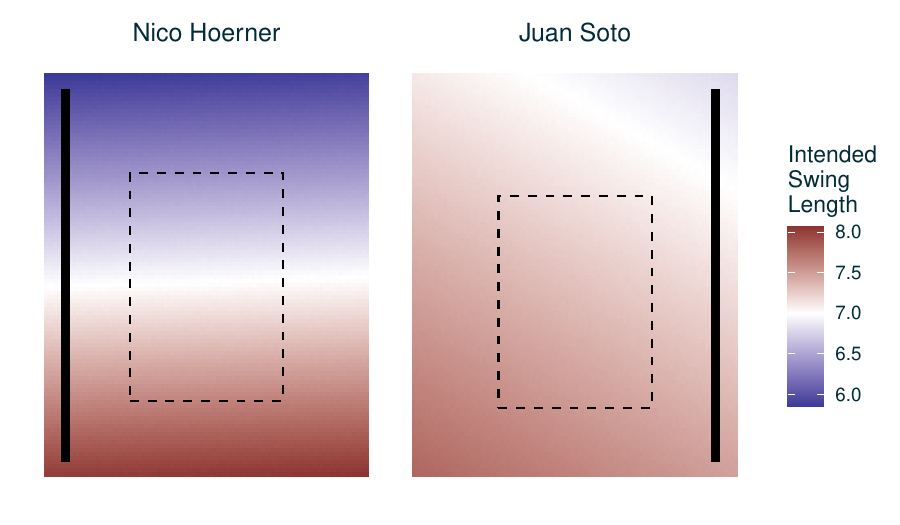}
        \caption{\it Expected swing length by pitch location for two sample batters: Nico Hoerner (left) and Juan Soto (right), viewed from behind home plate. The thick vertical bars show the side of the plate on which the batter stands, and the dashed rectangle shows the strike zone, which depends on the height and stance of the batter. The color shows the predicted swing length  from model (\ref{eqn:intention-swing-length}) assuming a 0-0 count.}
        \label{fig:adaptation}
      \end{figure}

    While Section \ref{sec:results-value} illustrated the tradeoff involved with how batters vary their \textit{approach}, there are of course other sources of variation in bat speed and swing length. For example, in Section \ref{sec:methods-intention} we define \textit{adaptation} as the extent to which a batter's swing tracking metrics co-vary with pitch location. As captured by $\sigma_b^X$ and $\sigma_b^Z$ in Table \ref{tab:intention-variances} and the posterior means of the batter-level random effects in Figure \ref{fig:intent-re}, we observe differences between batters in their swing \textit{adaptation} as the pitch location varies horizontally and vertically. To demonstrate the variation, Figure \ref{fig:adaptation} displays the expected swing length from the intention model by pitch location for two sample batters, Nico Hoerner and Juan Soto (assuming a 0-0 count). Hoerner displays a clear relationship with the vertical location with shorter intended swings on average for pitches that are located higher in the zone. In comparison, Soto displays a fairly consistent swing length throughout the zone, with an uptick in length for low and away pitches, versus a shortening of the swing for high and inside pitches. Understanding why batters display such variation in their \textit{adaptation} is left for future work, and could be explored with swing path measurements describing the tilt and attack angle that were released during the 2025 MLB season (\cite{petriello_4_2025}).
    
\begin{figure}
        \centering
        \includegraphics[width = 0.8\textwidth]{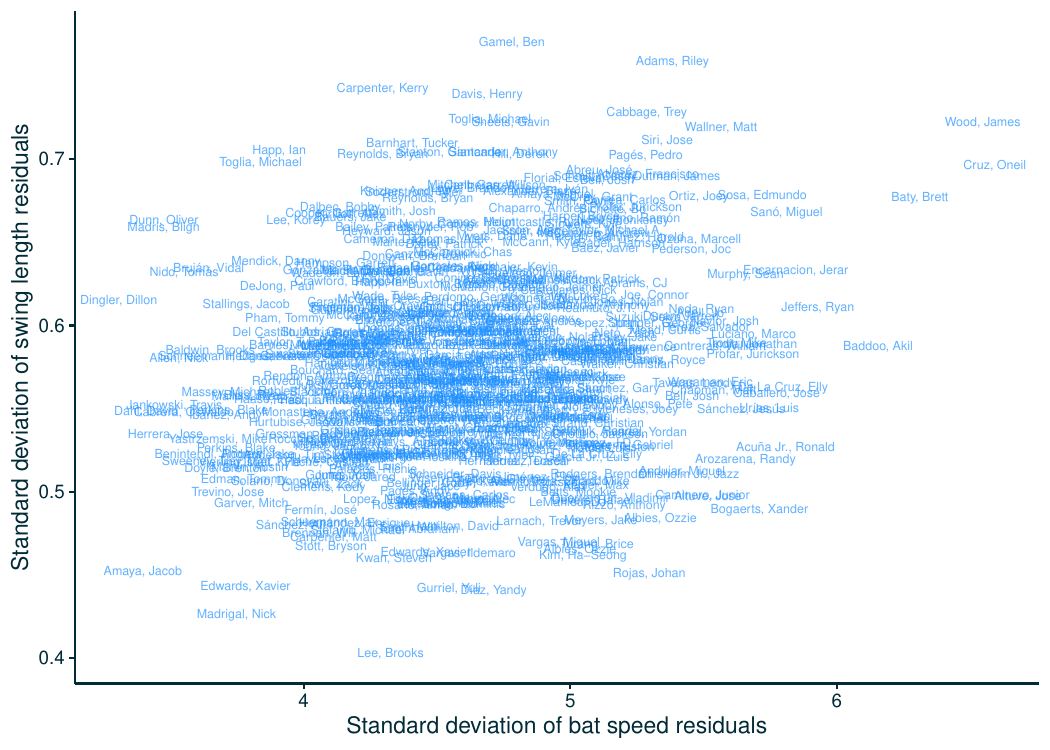}
        \caption{\it Joint distribution of the standard deviations of the swing length (y-axis) and bat speed (x-axis) residuals for all swings (not just squared up on primary fastballs) relative to the intention model predictions. Each batter is displayed by their name, and only batters with at least 25 squared-up swings are displayed.}
        \label{fig:resid-sd}
      \end{figure}

      While our definitions for \textit{approach} and \textit{adaptation} are based on the random effects of the intention model, there is additional unexplained variation remaining. Figure \ref{fig:resid-sd} displays the standard deviation of the residuals for all swings (not just squared up on primary fastballs) relative to the intention model predictions. We observe clear differences in the remaining variation between batters. For example, Oneil Cruz displays higher residual standard deviations for both bat speed and swing length while Yandy Diaz deviates relatively less from his intention model swing length. We hypothesize that the variation of these residuals is largely due to batter \textit{timing}. Further exploration of this is left for future work, which can include exploration of measurements  for the ``intercept point'', indicating where the bat makes contact with the ball with respect to the position of the batter's stance and home plate, which was released by MLB during the 2025 season.
      
    \subsection{Future Work}
    \label{sec:future-work}

      We limited our analysis of the run value of approach to its effect on a batter of average skill, but it is possible that the results in Section \ref{sec:results-value} could differ depending on the profile of the batter. For example, the value of a specific approach could be different for high-contact, low-power batters versus low-contact, high-power batters. Future work could estimate personalized effects of batter approach based on the batter's strengths and weaknesses. Additionally, the estimated run value was based on linear weights, which ignore the base-out situation. With a runner on third base and fewer than two outs, the value of putting the ball in play (relative to non-ball-in-play outcomes) increases, so presumably the tradeoff of increased contact for decreased power is more valuable in this situation. Future work could estimate this effect and evaluate the extent to which batters tailor their approach to the base-out situation.

      Additional releases of metrics from MLB could help with future work on swing tracking data. Bat speed and swing length, measured at contact, provide only a hint at the full swing path and need to be interpreted carefully. In May 2025, MLB released three new pitch-by-pitch level metrics describing the orientation and movement of the bat at contact: attack angle, attack direction and swing tilt (\cite{petriello_4_2025}). MLB has hinted that more metrics are in the works, such as contact depth and miss distance (\cite{petriello_everything_2024}). These two metrics would be particularly interesting because contact depth would help contextualize swing length and bat speed, and miss distance would provide significantly more precise data for batter and pitcher evaluation as compared with binary contact/miss outcomes.

  \subsection*{Acknowledgments}

    The authors thank Noah Woodward and Nick Wan for early stage discussions and feedback on this work.

  \subsection*{Code availability}

  All code related to this paper is available at \url{https://github.com/saberpowers/swinging-fast-and-slow}.

  \printbibliography

\end{document}